\documentstyle[12pt]{article}
\newtheorem{theo}{Theorem}
\title{On quasi-exactly solvable matrix models}
\author{R.Z.~Zhdanov \\ \small Institute of Mathematics,\\
\small 3 Tereshchenkivska Street,
252004 Kiev, Ukraine\thanks{e-mail: rzhdanov@apmat.freenet.kiev.ua}}
\date{}
\begin{document}
\maketitle
\begin{abstract}
An efficient procedure for constructing quasi-exactly solvable
matrix models is suggested. It is based on the fact that the
representation spaces of representations of the algebra $sl(2, {\bf R})$
within the class of first-order matrix differential operators
contain finite dimensional invariant subspaces.
\end{abstract}

The Lie-algebraic approach to constructing quasi-exactly solvable
one-dimensional stationary Schr\"odinger equations as suggested
by Shifman and Turbiner \cite{tur,shi} is based on the properties
of the representations of the algebra $sl(2,{\bf R})$
\begin{equation}
\label{1}
[Q_0,\ Q_{\pm}]=\pm Q_{\pm},\quad [Q_-,\ Q_+]=2Q_0
\end{equation}
by first-order differential operators. Namely, the approach in
question utilizes the fact that the representation space of the algebra
$sl(2,{\bf R})$ having the basis elements
\begin{equation}
\label{lvf}
Q_-={d\over dx},\quad Q_0=x{d\over dx} - n,\quad
Q_+=x^2{d\over dx} - 2nx,
\end{equation}
where $n$ is an arbitrary natural number, has an $(n+1)$-dimensional
invariant subspace. Its basis is formed by the polynomials in $x$ of
the order not higher than $n$ (for further details see the monograph
by Ushveridze \cite{ush} and references therein).

The aim of the present paper is to extend the above Lie-algebraic
approach in order to make it applicable to analyzing eigenvalue
problems for matrix differential operators. The key idea is
that the basis elements of the algebra $sl(2, {\bf R})$ are
searched for within the class of matrix differential operators
\begin{equation}
\label{matrix}
Q=\xi(x){d\over dx} + \eta(x),
\end{equation}
where $\xi(x), \eta(x)$ are some matrix-valued functions of
the corresponding dimension. Furthermore, the representation
space of $sl(2, {\bf R})$ must contain a finite-dimensional
subspace. Provided these requirements are met, a quasi-exactly
solvable matrix model is obtained by composing a linear
combination of the basis elements of the algebra $sl(2, {\bf R})$
with constant matrix coefficients.

Thus, to get a quasi-exactly solvable matrix model we need to
solve two intermediate problems
\begin{enumerate}
  \item[{1)}] {to solve the relations (\ref{1}) within the
               class (\ref{matrix}),}
  \item[{2)}] {to pick out from the set of thus obtained realizations
               of the algebra $sl(2, {\rm C})$ those ones whose
               representation space contains a finite-dimensional
               invariant subspace.}
\end{enumerate}

In a sequel we will restrict our considerations to the case when
$\xi(x)$ is a scalar multiple of the unit matrix. Given this
restriction, a simple computation yields that any representation of
$sl(2, {\bf R})$ within the class of operators (\ref{matrix}) is
equivalent to the following one:
\begin{equation}
\label{basis}
Q_-={d\over dx},\quad Q_0=x{d\over dx} + {A},\quad
Q_+=x^2{d\over dx} + 2x{A} + {B},
\end{equation}
where ${A},{B}$ are constant $N\times N$ matrices satisfying the relation
\begin{equation}
\label{2}
[{A},\ {B}]={B}.
\end{equation}

Next, we have to investigate under which circumstances the
representation space of the algebra (\ref{basis}) has a
finite-dimensional invariant subspace ${\cal I}_n$ with basis
elements
\[
{\vec f}_i(x)=\sum_{j=1}^N\, F_{ij}(x){\vec e}_j,\quad i=1,\ldots,n.
\]
Here $\vec e_1,\ldots, \vec e_N$ is the orthonormal basis of
the space ${\bf R}^N$. It occurs that the components of the functions
$\vec f_i$ are necessarily polynomials in $x$ of the order not higher
than $n-1$.
\begin{theo}
Let the functions $\vec f_1(x),\ldots,\vec f_n(x)$ form the basis of
the invariant subspace ${\cal I}_n$ of the representation space of
the Lie algebra $\langle {d\over dx}, x{d \over dx} + A\rangle$,
where $A$ is a  constant $N\times N$ matrix. Then
\begin{equation}
\label{lemma1}
{d^n\vec f_i(x)\over dx^n}=\vec 0,\quad i=1,\ldots,n.
\end{equation}
\end{theo}

We will give a sketch of the proof for the case, when $N=2$.
The requirement that a representation space of a Lie algebra under
study contains a finite-dimensional invariant subspace means
that the following relations hold
\begin{eqnarray}
&&{d\over dx} {\vec f}_i=\sum_{j=1}^n\, \Lambda_{ij}{\vec f}_j,
\label{form1}\\
&&\left(x{d\over dx} + A\right ) {\vec f}_i=\sum_{j=1}^n\,
L_{ij}{\vec f}_j,\label{form2}
\end{eqnarray}
where $\Lambda_{ij}, L_{ij}$ are arbitrary complex constants,
$i,j =1,\ldots, n$.

Solving (\ref{form1}) yields
\begin{equation}
\label{exp}
{\vec f}_i(x)=\sum_{j=1}^N\, \sum_{k=1}^n\,\left({\rm e}^{\Lambda x}
\right)_{ik} C_{kj}{\vec e}_j,
\end{equation}
where $C_{ij},\ i=1,\ldots,n,\ j=1,\ldots,N$ are arbitrary complex
constants and the symbol $(A)_{ij}$ stands for the $(i,j)$th entry of
the matrix $A$.

Next, from the requirement (\ref{form2}) we get
\begin{equation}
\label{con1}
x\Lambda C + C A ={\rm e}^{-\Lambda x}\, L\, {\rm e}^{\Lambda x} C.
\end{equation}
Making use of the Cambell-Hausdorff formula and equating
coefficients of the powers of $x$ give the following
infinite set of algebraic equations for unknown
matrices $L, \Lambda, C, A$:
\begin{eqnarray}
&& LC = CA,\label{eq1}\\
&& [L,\ \Lambda]C = \Lambda C,\label{eq2}\\
&& \{L,\ \Lambda\}^i\,C=0,\quad i\ge 2,\label{eq3}
\end{eqnarray}
where
\[
\{L,\ \Lambda\}^0=L,\quad
\{L,\ \Lambda\}^i=[\{L,\ \Lambda\}^{i-1},\ \Lambda],
\quad i\ge 1.
\]

Choosing the basis vectors $\vec e_1, \vec e_2$ in an appropriate
way, we can  transform the constant matrix $A$ to the Jordan form.
There are two possibilities
\begin{eqnarray}
A&=&\left(\begin{array}{cc} \lambda& 1\\ 0&\lambda\end{array} \right),
\label{mat1}\\
A&=&\left(\begin{array}{cc} \lambda_1& 0\\ 0&\lambda_2\end{array}\right).
\label{mat2}
\end{eqnarray}
Here $\lambda, \lambda_1, \lambda_2$ are arbitrary constants.
\vspace{2mm}

\noindent
{\bf Case 1.} Let the matrix $A$ be of the form (\ref{mat1}). If we denote
the first and the second columns of the $2\times 2$ matrix $C$ as $\vec C_1$
and $\vec C_2$, then equations (\ref{eq1}) can be rewritten to become
\[
L \vec C_1 = \lambda \vec C_1,\quad
L \vec C_2 = \vec C_1 + \lambda \vec C_2.
\]
Next, using equations (\ref{eq2}), (\ref{eq3}) we obtain the following
relations
\begin{eqnarray}
L\Lambda^j\vec C_1 = \lambda_j\Lambda^j\vec C_1,\label{c1}\\
L\Lambda^j\vec C_2 = \Lambda^j\vec C_1 + \lambda_j\Lambda^j\vec C_2.\label{c2}
\end{eqnarray}
Here $\lambda_0=\lambda,\ \lambda_{j+1}=\lambda_j + 1$,\ $j=0,1,\ldots$.

It follows from (\ref{c1}) that $\vec a_{i+1} = \Lambda^i\vec C_1,\ i\ge 0$
are eigenvectors of the $n\times n$ matrix $L$ corresponding to eigenvalues
$\lambda_i=\lambda + i$ and, what is more, since these eigenvectors correspond
to distinct eigenvalues, they are linearly independent. As there are at most
$n$ linearly independent eigenvectors of the matrix $L$, the relation
\begin{equation}
\label{zero1}
\Lambda^m \vec C_1 =\vec 0
\end{equation}
with some $m\le n$ holds true.

Combining (\ref{c2}) and (\ref{zero1}) yields
\begin{equation}
\label{eigen}
L\Lambda^{m+i} \vec C_1 =\lambda_{m+i}\Lambda^{m+i}\vec C_2,
\quad i\ge 0.
\end{equation}
Hence, we conclude that the vectors $\vec a_{m+i} = \Lambda^{m+i}\vec
C_2,\ i\ge 0$ are eigenvectors of the matrix $L$ forming
together with the vectors $\vec a_1,\ldots,\vec a_m$ the system
of its linearly independent eigenvectors. As an $n\times n$ matrix
has at most $n$ linearly independent eigenvectors,
the relation
\begin{equation}
\label{zero2}
\Lambda^n \vec C_2 =\vec 0
\end{equation}
holds true.

In view of (\ref{zero1}), (\ref{zero2}) the matrix $C=(\vec C_1,
\vec C_2)$ satisfy the following matrix equation
\[
\Lambda^n\, C=0.
\]
Due to this fact, (\ref{exp}) reads as
\begin{equation}
\label{exp1}
{\vec f}_i(x)=\sum_{j=1}^N\, \sum_{k=1}^n\,\left(1 + x\Lambda +
\frac{x^2}{2!}\Lambda^2 + \cdots + \frac{x^{n-1}}{(n-1)!}
\Lambda^{n-1} \right)_{ik}
C_{kj}{\vec e}_j.
\end{equation}
Thus, the components of the vectors $\vec f_j$ are polynomials
of the order not higher than $n-1$, which is the same as what
was to be proved.
\vspace{2mm}

\noindent
{\bf Case 2.}\ We turn now to the case when the matrix $A$ is given
by (\ref{mat2}). Denoting the first and the second columns of the
$2\times 2$ matrix $C$ as $\vec C_1$ and $\vec C_2$, we rewrite
equations (\ref{eq1}) as follows
\[
L \vec C_1 = \lambda_1 \vec C_1,\quad
L \vec C_2 = \lambda_2 \vec C_2.
\]
Next, using equations (\ref{eq2}), (\ref{eq3}) we obtain the relations:
\begin{eqnarray}
L\Lambda^j\vec C_1 = \alpha_j\Lambda^j\vec C_1,\label{c1a}\\
L\Lambda^j\vec C_2 = \beta_j\Lambda^j\vec C_2,\label{c2a}
\end{eqnarray}
where
\begin{eqnarray*}
&&\alpha_0=\lambda_1,\quad \alpha_{j+1}=\alpha_j + 1,\quad
j=0,1,\ldots,\\
&&\beta_0=\lambda_2,\quad \beta_{j+1}=\beta_j + 1,
\quad j=0,1,\ldots.
\end{eqnarray*}

Thus, the vectors $\vec e_{i+1}=\Lambda^i\vec C_1,\ i=0,1,\ldots$ are
eigenvectors of the $n\times n$ matrix $L$ with eigenvalues
$\alpha_i=\lambda_1 + i,\ i=0,1,\ldots$. As these eigenvalues are
distinct, the vectors $\vec e_i$ are linearly independent.
Taking into account that an $n\times n$ matrix can have at
most $n$ linearly independent eigenvectors we conclude that
$\Lambda^n\vec C_1=\vec 0$. Similarly, we get the relation
$\Lambda^n \vec C_2 = \vec 0$. Hence it follows that the
expression (\ref{exp}) takes the form (\ref{exp1}), which
is the same as what was to be proved. $\bullet$
\vspace{2mm}

Consequently, the most general finite-dimensional invariant subspace
of the representation space of the algebra (\ref{basis}) is spanned
by vectors whose components are finite-order polynomials in $x$.
We postpone a detailed study of these representations with
arbitrary $N$ for future publications and concentrate on the
case $N=2$. There are two families of inequivalent finite dimensional
representations of the algebra $sl(2, {\bf R})$ having the
basis elements (\ref{basis})
\begin{eqnarray}
&{\rm I.}& A=\left(\begin{array}{cc} -\frac{n}{2} & 0\\
                               0 & -\frac{m}{2}
      \end{array}\right),
\quad B=\left(\begin{array}{cc} 0 & 0\\
                                0 & 0\end{array}\right),
\label{rep1}\\[4mm]
&{\rm II.}& A=\left(\begin{array}{cc} -\frac{n}{2} & 0\\
                               0 & \frac{2-n}{2} \end{array}\right),
\quad B=\left(\begin{array}{cc} 0 & 0\\
                                -1 & 0\end{array}\right).
\label{rep2}
\end{eqnarray}
Here $n, m$ are abritrary natural numbers with $n\ge m$.

Representations of the form (\ref{basis}), (\ref{rep1}) are the
direct sums of two irreducible representations realized on the
representation spaces
\[
{\cal R}_1=\langle \vec e_1, x\vec e_1,\ldots,
x^n\vec e_1\rangle,\quad
{\cal R}_2=\langle \vec e_2, x\vec e_2,\ldots, x^m\vec e_2\rangle,
\]
where $\vec e_1=(1,0)^{\rm T},\ \vec e_2=(0,1)^{\rm T}$.

Next, representations (\ref{basis}), (\ref{rep2}) are also the
direct sums of two irreducible representations realized on the
representation spaces
\begin{eqnarray*}
{\cal R}_1&=&\langle n\vec e_1,\ldots,nx^j\vec e_1 +
jx^{j-1}\vec e_2,\ldots,nx^n\vec e_1+nx^{n-1}\vec e_2\rangle,\\
{\cal R}_2&=&\langle \vec e_2, x\vec e_2,\ldots,
x^{n-2}\vec e_2\rangle.
\end{eqnarray*}

We will finish the paper with an example of utilizing
the above results for obtaining an exactly solvable
two-component Dirac-type equation, which is one of the two
differential equations composing the Lax pair for the
cubic Schr\"odinger equation (see, e.g. \cite{abl,zah}).
Consider the following two-component matrix model:
\begin{equation}
\label{exa1}
{\cal H}\vec w\equiv i(ax\sigma_2 + b\sigma_1)
{\displaystyle d\vec w\over \displaystyle dx} +
\left(c_1\sigma_1 + \left(c_2
+ ia\right)\sigma_2\right)\vec w = \lambda\vec w,
\end{equation}
where $a, b, c_1, c_2$ are arbitrary real parameters with
$ab\ne 0$ and $\sigma_1, \sigma_2$ are $2\times 2$
Pauli matrices. As
\[
{\cal H}= ib\sigma_1Q_- + ia\sigma_2Q_0 +
c_1\sigma_1 + c_2\sigma_2,
\]
where $Q_-, Q_+$ are given by (\ref{basis}) with $A=1$,
the operator ${\cal H}$ transforms the $(2n+2)$-dimensional vector
space
\[
\vec f_{j}(x)=x^{j-1}\vec e_1,\quad \vec f_{n+j+1}(x)=x^{j-1}\vec e_2,
\quad j=1,\ldots,n+1
\]
into itself. This means that there exists the constant
$(2n+2)\times (2n+2)$ matrix $H$ such that
\[
{\cal H}\vec f_i=\sum\limits_{j=1}^{2n+2}\,H_{ij}\vec f_j,\quad
i=1,\ldots,2n+2.
\]
Hence it immediately follows that the vector-function
\[
\vec \psi(x) = \sum\limits_{j=1}^{2n+2}\,\alpha_j\vec f_j(x)
\]
is the solution of the system of ordinary differential equations
(\ref{exa1}) with $\lambda=\lambda_0$, provided $(\alpha_1,\ldots,
\alpha_{2n+2})$ is an eigenvector of the matrix $H$ with the
eigenvalue $\lambda_0$.

Making a transformation
\begin{eqnarray*}
x&=&\frac{b}{a}\sinh(ay),\\
\vec w(x)&=& (\cosh(ay))^{1/2}\exp\left\{
-\frac{i}{a}\left(c_1\arctan\sinh(ay) + c_2\ln\cosh(ay)\right)\right\}\\
&&\times\exp\{-i\sigma_3\arctan\sinh(ay)\}\,\vec \psi(y)
\end{eqnarray*}
we reduce (\ref{exa1}) to the Dirac-type
equation
\begin{equation}
\label{exa2}
i\sigma_1{d\vec \psi\over dy} + \sigma_2 V(y)\vec \psi(y)=
\lambda \vec\psi,
\end{equation}
where
\[
V(y)={a^2c_2-b^2c_1\sinh(ay)\over ab\cosh(ay)}
\]
is the well-known hyperbolic P\"oschel-Teller potential. It
has exact solutions of the form
\begin{eqnarray*}
\psi(y)&=&(\cosh(ay))^{-1/2}\exp\left\{
\frac{i}{a}\left(c_1\arctan\sinh(ay) + c_2\ln\cosh(ay)\right)\right\}\\
&&\times\exp\{i\arctan\sinh(ay)\sigma_3\}
\sum\limits_{j=1}^{2n+2}\,\alpha_j\vec f_j\left(\frac{b}{a}
\sinh(ay)\right).
\end{eqnarray*}

As the potential $V$ does not depend explicitly on $n$, the
order of the polynomials $P_n, Q_n$ may be arbitrarily
large. This means that the Dirac equation (\ref{exa2}) with the
hyperbolic P\"oschel-Teller potential is {\em exactly-solvable}.

In \cite{zhd} we suggest an alternative approach to construction
of quasi-exactly solvable stationary Schr\"odinger equations
based on their conditional symmetry. We believe that a similar
idea should work for matrix models as well. It is intended to
devote one of the future publications to a comparison of
the conditional symmetry and Lie-algebraic approaches to
constructing quasi-exactly solvable Dirac-type equations
(\ref{exa2}).

\section*{Acknowledgements}
It is a pleasure for the author to thank Anatolii Nikitin for
a fruitful discussion of some aspects of this work and Alex
Ushveridze who encourage him to present the above material
as a paper.

\end{document}